\documentclass[preprint,preprintnumbers,amsmath,amssymb]{revtex4}
\usepackage{graphicx} 
\usepackage{setspace} 
\usepackage{dcolumn} 
\usepackage{bm} 
\usepackage{natbib}
\usepackage{hyperref}
\hypersetup{colorlinks=true,linkcolor=blue,citecolor=blue,urlcolor=blue}
\usepackage{epsfig}
\usepackage{amsmath}
\usepackage{epsf,graphics,graphicx}
\begin{document}
\title{Thermodynamic origin of entropy-assisted stabilization in multicomponent rock-salt oxides}

\author{Supriya Ghosal$^a$}
\author{Swapan K. Pati$^a$}
\author{Ashutosh Kumar$^{b,}$\footnote{Email: ashutosh@iitbhilai.ac.in}}
\affiliation{$^a$Theoretical Sciences Unit, Jawaharlal Nehru Centre for Advanced Scientific Research, Bangalore, Karnataka 560064, India}
\affiliation{$^b$Functional Materials Laboratory, Department of Materials Science and Metallurgical Engineering, Indian Institute of Technology Bhilai, Chhattisgarh 491002, India}
\date{\today}

\begin{abstract}
High-entropy oxides are commonly associated with high configurational entropy ($\Delta S_{conf}\geq$ 1.61R) corresponding to five equimolar cations occupying a crystallographic sublattice. However, recent experimental observations indicate that medium-entropy compositions may also exhibit entropy-stabilized rock-salt phases, raising an important question regarding the minimum entropy required for phase stabilization. In this work, we employ a first-principles thermodynamic framework to investigate the stability of rock-salt oxides containing two to five principal cations components, analogous to (Ni$_{0.8}$Cu$_{0.2}$)O, (Ni$_{0.6}$Cu$_{0.2}$Zn$_{0.2}$)O, (Ni$_{0.4}$Cu$_{0.2}$Zn$_{0.2}$Co$_{0.2}$)O, and (Ni$_{0.2}$Cu$_{0.2}$Zn$_{0.2}$Co$_{0.2}$Mg$_{0.2}$)O. Density functional theory, MCSQS-based structural modeling, and finite-temperature Gibbs free-energy analysis are combined to quantify the roles of enthalpy mixing ($\Delta H_{mix}$), configurational ($\Delta S_{conf}$), vibrational ($\Delta S_{vib}$), electronic ($\Delta S_{elec}$), and magnetic ($\Delta S_{mag}$) contributions towards entropy change in governing phase stability. The results show that $\Delta S_{conf}$ alone is not a universal descriptor of phase stability. While the two-cation system is enthalpy-stabilized but, three-, four- and five-cation systems become thermodynamically stable at high temperatures due to entropy-driven reduction of the Gibbs free energy. These findings demonstrate that single-phase rock-salt oxides are not restricted to the conventional high-entropy limit and that medium-entropy compositions can also be stabilized under suitable thermodynamic conditions.\\
\end{abstract}
\maketitle
\section{Introduction}
High-entropy oxides (HEOx) have opened a new direction in the chemistry and physics of crystalline solids by extending the concept of multicomponent disorder from metallic systems to oxygen-mediated compounds. In these materials, several cations share a crystallographic sublattice in near-equimolar proportions, producing large chemical disorder within a still-crystalline framework. Unlike conventional substitutional oxides, where disorder is typically treated as a perturbation to a parent lattice, HEOx represent a regime in which disorder itself becomes a defining structural parameter. This has enabled access to unusual combinations of phase stability and functionality across a wide range of oxide systems.\\
The field of HEOx was catalyzed by the discovery of single-phase entropy stabilized oxide: (Mg,Co,Ni,Cu,Zn)O, in which five cations were incorporated into one cationic sublattice and stabilized as a homogeneous solid solution at elevated temperature.\cite{rost2015entropy} That discovery established a central idea in the field: a configurational entropy $\Delta S_{conf}$ $\geq$ 1.6R can offset unfavorable enthalpic contributions and stabilize compositionally complex oxide phases that would otherwise decompose into simpler constituents. Since then, entropy-assisted stabilization has been explored in several oxide families, including rock-salt, fluorite\cite{kumar2023novel}, pyrochlore\cite{vayer2021new}, and perovskite\cite{spiridigliozzi2024design} structures, indicating that multicomponent cation disorder can serve as a general route toward structurally and functionally diverse oxide materials.\\
Fracchia et al.\cite{fracchia2022configurational} demonstrated that rock-salt oxides with intermediate configurational entropy can exhibit entropy-stabilized behavior, consistent with earlier studies showing that entropic stabilization is not restricted to systems containing five or more components\cite{2021t}. 
Subsequent experimental studies of the prototypical (Mg,Co,Ni,Cu,Zn)O system have further demonstrated that its phase stability cannot be described solely by the magnitude of configurational entropy. Temperature-dependent synchrotron diffraction has revealed structural distortions and multistep segregation pathways involving guggenite and tenorite phases,\cite{coduri} while reversible entropy-driven phase transformations and strong grain-size effects have also been observed experimentally.\cite{dupay} Furthermore, systematic experimental and first-principles investigations have shown that the stability contribution of individual cations can differ substantially and depends on the competing phase chemistry and synthesis conditions.\cite{Acta} These observations emphasize that configurational entropy is an important but not sufficient descriptor of phase stability. This observation calls for a basic thermodynamic question that remains unresolved: how much $\Delta S_{conf}$ is actually required to stabilize a single-phase oxide? While the common operational definition of a high-entropy material is strictly mapped to an ideal equimolar five-component system ($\Delta S_{\mathrm{conf}} \approx 1.61R$), \cite{rost2015entropy, spurling2022entropy, aamlid2023understanding} relaxed thresholds (e.g., $\Delta S_{\mathrm{conf}} \geq 1.5R$) are widely accepted to accommodate non-equimolar deviations. Regardless of the exact threshold, a large configurational entropy alone does not necessarily prove entropy-stabilized phase formation. Also, It is important to distinguish between high-entropy oxides and entropy-stabilized oxides. A high-entropy oxide is usually classified based on compositional complexity and configurational entropy \cite{kumar2023magnetic, kumar2024thermoelectric}, whereas an entropy-stabilized oxide is one in which the single-phase structure is stabilized specifically because the entropy contribution overcomes the positive enthalpy penalty for mixing\cite{brahlek2022name,aamlid2023understanding}. Therefore, a large $\Delta S_{\mathrm{conf}}$ alone does not necessarily prove entropy-stabilized phase formation; rather, the stabilization must be evaluated from the temperature-dependent Gibbs free energy. This distinction is particularly important for medium-entropy oxides, where the configurational entropy is lower than the conventional high-entropy threshold but may still be sufficient to stabilize a single-phase oxide under suitable thermodynamic conditions.\\
More fundamentally, the stability of a multicomponent oxide may not be inferred from configurational entropy alone. The relevant thermodynamic quantity is the Gibbs free energy of mixing, in which the enthalpy of mixing competes with temperature-dependent entropy contributions. In addition to configurational entropy, both vibrational and electronic entropy may influence the stabilization of compositionally complex oxides, particularly at elevated temperature. Therefore, whether a given rock-salt oxide forms a stable single phase depends on the collective balance among these contributions rather than on a nominal entropy classification alone. Resolving this issue is essential not only for clarifying the physical meaning of “entropy stabilization,” but also for establishing predictive principles for the design of compositionally complex oxides.\\
Here, we develop a first-principles framework to elucidate the thermodynamic phase stability of rock-salt oxides containing two to five principal cation components. Recent descriptor-based studies have also shown that the formation of single-phase multicomponent oxides cannot be predicted solely from configurational entropy.\cite{sivak2025discovering, divilov2024disordered, manchon2025descriptors} These studies emphasize that parameters such as statistical variations in local mixing enthalpies (e.g., the entropy-forming ability), cation-size mismatch (tolerance factors), preferred binary oxide ground states, and local bonding environments must be evaluated collectively. In particular, recent thermodynamics-inspired studies have emphasized that enthalpic stability, cation–anion bonding characteristics, oxidation-state compatibility, and oxygen chemical potential can play decisive roles in determining the synthesizability and stability of rock-salt high-entropy oxides.\cite{almishal-nature} Thus, a comprehensive description of HEO stability requires consideration of both energetic and entropic contributions together with the relevant chemical environment and processing conditions. This motivates a more complete thermodynamic treatment in which $\Delta H_{\mathrm{mix}}$, $\Delta S_{\mathrm{conf}}$, $\Delta S_{\mathrm{vib}}$, $\Delta S_{\mathrm{elec}}$ and \textcolor{blue}{$\Delta S_{mag}$} are evaluated explicitly rather than assuming a fixed entropy threshold for phase stabilization. Recent DFT and MLIP studies have provided important insights into HEO energetics and stability.\cite{Acta,sivak2025discovering} Here, we focus on the temperature-dependent balance between mixing enthalpy and multiple entropy contributions to assess thermodynamic stability across different cation complexities. In this work, we go beyond configurational-entropy-based stability criteria by systematically examining the temperature-dependent competition between mixing enthalpy and configurational, vibrational, electronic and magnetic entropy contributions in rock-salt oxide solid solutions spanning two- to five-cation components. By comparing compositions with different degrees of chemical complexity within a common thermodynamic framework, we demonstrate that the magnitude of configurational entropy alone does not determine stability; rather, the balance between the enthalpic penalty and the total entropy contribution governs the temperature-dependent $\Delta G_{\mathrm{mix}}$.

%
%
%

%
%
%
\section{Computational Methods}

Theoretical analyses were carried out using the first-principles density functional theory (DFT) framework as implemented in the Vienna \textit{Ab-initio} Simulation Package (VASP)\cite{kresse1993ab,kresse1994ab,kresse1996efficiency}. In order to describe ion-electron interactions accurately, the projector augmented-wave (PAW) method was employed in conjunction with the generalized gradient approximation (GGA) in the form of the Perdew-Burke-Ernzerhof (PBE) exchange-correlation functional\cite{paw,GGA}. For structural optimizations, the unit cells were sampled using a $4\times4\times4$ Monkhorst-Pack (MP) $k$-point grid in combination with a plane-wave cutoff energy of $600\,\mathrm{eV}$\cite{mp}. The unit-cell structures were geometrically optimized until the residual force components converged below $10^{-3}\,\mathrm{eV/\AA}$, with a total energy convergence tolerance of $10^{-7}\,\mathrm{eV}$ for the self-consistent electronic loop. To account for on-site Coulomb interactions in the localized transition-metal 3d states, spin-polarized GGA+$U$
calculations were performed using the simplified rotationally invariant scheme of Dudarev et al.\cite{dudarev} In this approach, the correction is defined by the effective Hubbard parameter, $U_{eff}=U-J$, and was applied to the 3d orbitals of the transition-metal atoms.  The $U_{eff}$ values applied to the localized 3d orbitals were 6.4 eV for Ni, 4.0 eV for Cu, and 3.3 eV for Co as implemented in the previous reference.\cite{wang2006oxidation}

Depending on the experimentally reported structure of NiO\cite{extra} with fractional occupancy, random structures were generated using the special quasi-random structures (SQS) method. The SQS structures were generated by imposing chemical disorder on the cation sublattice without assigning a specific magnetic ordering. SQS formalism uses the mcsqs code as implemented in the Alloy Theoretic Automated Toolkit (ATAT) routines\cite{sqs,van2002alloy,van2009multicomponent,van2013efficient}. The SQS algorithm is designed to minimize the correlation difference between a finite cell and the corresponding infinite random system. Thus, SQSs provide a suitable platform to approximate near-randomness in the properties of solid-solution alloys. For all four structures included in the theoretical study, a $60$-atom supercell was used for the SQS structures with cutoff radii of 8 {\AA}, 7 {\AA}, and 6 {\AA} for pair, three-body, and four-body correlations, respectively.

The SQS approach was used to approximate a random cation distribution on the rock-salt sublattice. However, it should be noted that the ideal configurational entropy expression assumes complete random mixing. In real multicomponent oxides, local chemical preferences and short-range ordering may reduce the effective configurational entropy. Therefore, the SQS-based results should be interpreted as a thermodynamic approximation to the disordered state.

To treat magnetic elements and their magnetic ground states accurately, \textcolor{blue}{spin-polarized calculations were subsequently performed using antiferromagnetic-like initial moments based on the reference magnetic oxides, and the moments were allowed to relax self-consistently during structural optimization.} For NiO and CoO, antiferromagnetic ordering was used in accordance with their known ground states. The magnetic moments were initialized on transition-metal cations, and the final converged magnetic moments were checked after structural relaxation. \textcolor{blue}{Except MO-14, all other MO systems converged to states with non-zero net magnetization.}


To assess phase stability, the change in Gibbs free energy of mixing, $\Delta G_{\mathrm{mix}}$, was calculated using the following relation\cite{ex1}:
\begin{equation}
\Delta G_{\mathrm{mix}}
=
\Delta H_{\mathrm{mix}}
-
T\Delta S_{\mathrm{mix}} .
\end{equation}
Here, $\Delta H_{\mathrm{mix}}$ and $\Delta S_{\mathrm{mix}}$ represent the enthalpy change and entropy change of mixing, respectively. For a consistent thermodynamic comparison, all energetic and entropic quantities were normalized on a per-atom basis. The mixing enthalpy was evaluated with respect to the thermodynamically stable binary oxide reference phases using the same computational settings. The total entropy of mixing was considered as the sum of configurational, vibrational, electronic, and magnetic entropy contributions:
\begin{equation}
\Delta S_{\mathrm{mix}}
= \Delta S_{\mathrm{conf}}
+ \Delta S_{\mathrm{vib}}
+ \Delta S_{\mathrm{elec}}
+ \Delta S_{\mathrm{mag}}.
\end{equation}
Therefore, the temperature-dependent phase stability was assessed from the total Gibbs free energy of mixing rather than from configurational entropy alone. Accordingly, the calculated $\Delta G_{\mathrm{mix}}$ values are used here as thermodynamic indicators relative to the chosen references, rather than as definitive phase-boundary or synthesis predictions.

The mixing enthalpy, $\Delta H_{\mathrm{mix}}$, was calculated using the following expression for a two-cation random solid solution decomposing into two one-cation oxides:
\begin{equation}
\Delta H_{\mathrm{mix}} =\;
E_{\mathrm{DFT}}\!\left[(A_mA'_n)O_{m+n}\right]
- mE_{\mathrm{DFT}}\!\left[AO\right]
- nE_{\mathrm{DFT}}\!\left[A'O\right] .
\end{equation}
Here, $E_{\mathrm{DFT}}\!\left[(A_mA'_n)O_{m+n}\right]$, $E_{\mathrm{DFT}}\!\left[AO\right]$, and $E_{\mathrm{DFT}}\!\left[A'O\right]$ represent the ground-state total energies of the two-cation random solid solution $(A_mA'_n)O_{m+n}$ and the corresponding one-cation oxides $AO$ and $A'O$, respectively. In a similar manner, the mixing enthalpy for a five-cation SQS structure was also calculated. The positive or negative values of $\Delta H_{\mathrm{mix}}$ reported in this manuscript should be interpreted only with respect to the selected binary-oxide references. They do not establish whether the multicomponent phase is stable against every possible decomposition pathway.

The mixing entropy includes contributions from configurational entropy, $S_{\mathrm{conf}}$, vibrational entropy, $S_{\mathrm{vib}}$, and electronic entropy, $S_{\mathrm{elec}}$\cite{ex1}. The change in mixing entropy represents the difference in entropy between the random solid-solution phase and the constituent binary oxides. The configurational entropy of the random solid solutions was calculated using:
\begin{equation}
S_{\mathrm{conf}}
=
-R\sum_{i=1}^{n} x_i \ln x_i ,
\end{equation}
where $x_i$ is the molar fraction of the $i$th cation on the cationic sublattice.

To calculate the vibrational entropy, phonon calculations were performed for the supercell. Based on the finite-displacement approach, a large number of displaced structures were generated using the Phonopy code\cite{t2}. The second-order interatomic force constants were obtained from static DFT calculations performed on multiple displaced configurations generated within a $2\times2\times2$ supercell. The dynamical matrix was constructed for different $q$-vectors in the Brillouin zone using the force-constant matrix. By solving the dynamical matrix, the eigenvalues of phonon frequencies and eigenvectors of phonon modes were obtained using the Phonopy package. The vibrational entropy was calculated by summing over all phonon modes using the following expression\cite{togo2015first}:
\begin{equation}
S_{\mathrm{vib}} =\;
\frac{1}{2T}\sum_{q,b}
\hbar\omega(q,b)
\coth\!\left[
\frac{\hbar\omega(q,b)}{2k_{\mathrm{B}}T}
\right]  
- k_{\mathrm{B}}\sum_{q,b}
\ln\!\left[
2\sinh\!\left(
\frac{\hbar\omega(q,b)}{2k_{\mathrm{B}}T}
\right)
\right] .
\end{equation}
Here, $\hbar$, $\omega(q,b)$, $k_{\mathrm{B}}$, and $T$ represent the reduced Planck constant, phonon frequency for wave vector $q$ and branch $b$, Boltzmann constant, and temperature, respectively.

The electronic contribution to entropy, $S_{\mathrm{elec}}$, was calculated from the electronic density of states by considering the thermal excitation of electrons across the Fermi level\cite{kresse1996efficiency}. The electronic entropy was evaluated using:
\begin{equation}
S_{\mathrm{elec}} =\;
-k_{\mathrm{B}}\int N(\varepsilon)
\Big[
f(\varepsilon)\ln f(\varepsilon) +
\left\{1-f(\varepsilon)\right\}
\ln\left\{1-f(\varepsilon)\right\}
\Big]\,d\varepsilon .
\end{equation}
Here, $k_{\mathrm{B}}$, $N(\varepsilon)$, and $f(\varepsilon)$ represent Boltzmann's constant, the electronic density of states, and the Fermi-Dirac distribution function, respectively. When finite-temperature electronic calculations were performed, the total electronic entropy was calculated as a time average over the equilibrated simulation trajectory:
\begin{equation}
S_{\mathrm{elec}}
=
\left\langle S_{\mathrm{elec}}(t)\right\rangle_t .
\end{equation}
The magnetic contribution to entropy, $S_{\mathrm{mag}}$ was evaluated within a mean-field approximation as employed in the previous theoretical reports of high entropy alloys \cite{k1,k2}. 
\begin{equation}
    S_{mag} = k_B \sum_i c_i ln (1+|M_i|)
\end{equation}
Here $k_B$, $c_i$, and $M_i$ represent the Boltzmann constant, the fractional concentration of atomic species $i$ and the corresponding DFT-calculated local magnetic moment (in $\mu_B$). This expression is based on the Boltzmann relation, $S = k_B ln \Omega$, by approximating the number of accessible magnetic states associated with species $\Omega_i=1+|M_i|$. The absolute value ensures that the entropy depends on the magnitude of the moment rather than its orientation or sign, so oppositely aligned moments contribute equally. This local-moment formulation assigns an effective magnetic-state degeneracy of $1+|M_i|$ to each constituent and compositionally averages its entropic contribution.
The change in entropy was then obtained as the difference between the entropy of the random solid-solution phase and the weighted entropy contributions of the corresponding constituent binary oxides. In the present work, thermal expansion, anharmonic effects and magnetic entropy contributions are not explicitly included. Therefore, the calculated Gibbs-energy changes and stabilization temperatures should be regarded as approximate estimates within the adopted computational framework.\\ 
\section{Results and Discussion}
\subsection{Mixing enthalpy of solid solutions}
\begin{figure*}
    \centering
    \includegraphics[width=15cm]{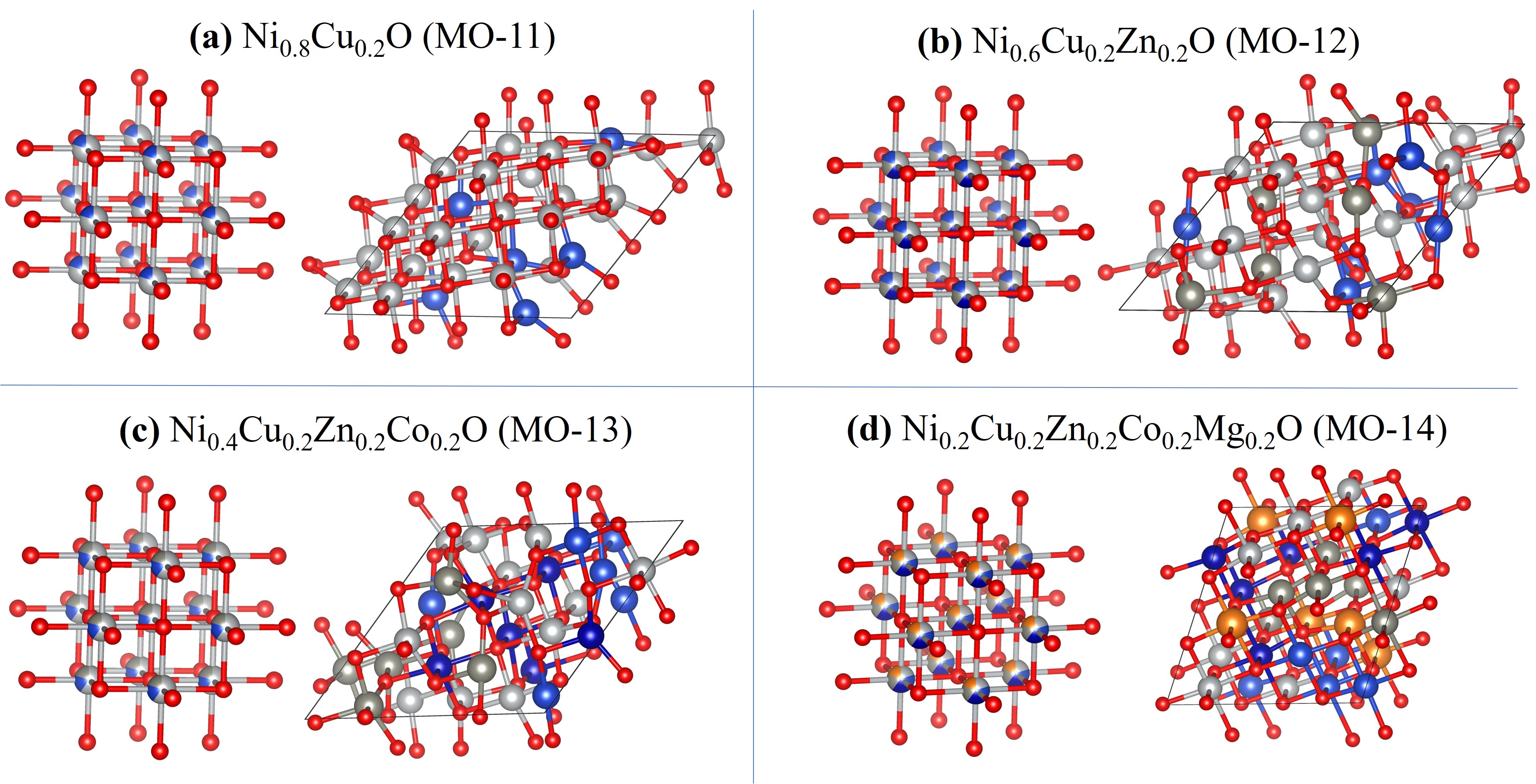}
    \caption{Structural geometry of  (a) Ni$_{0.8}$Cu$_{0.2}$O (MO-11), (b) Ni$_{0.6}$Cu$_{0.2}$Zn$_{0.2}$O (MO-12), (c) Ni$_{0.4}$Cu$_{0.2}$Zn$_{0.2}$Co$_{0.2}$O (MO-13), (d) Ni$_{0.2}$Cu$_{0.2}$Zn$_{0.2}$Co$_{0.2}$Mg$_{0.2}$O (MO-14) oxide systems. In each box, left panel and right panel showcase ordered crystalline and quasi-random disordered structures respectively.}
    \label{fig:str}
\end{figure*}
Theoretically, by means of first-principles DFT, ab-initio molecular dynamics (AIMD) and phonon calculations, thermodynamic phases of SQSs are analyzed. For theoretical study, we have considered 4 oxide systems i.e. MO-11 (Ni$_{0.8}$Cu$_{0.2}$O), MO-12 (Ni$_{0.6}$Cu$_{0.2}$Zn$_{0.2}$O), MO-13 (Ni$_{0.4}$Cu$_{0.2}$Zn$_{0.2}$Co$_{0.2}$O), MO-14
(Ni$_{0.2}$Cu$_{0.2}$Zn$_{0.2}$Co$_{0.2}$Mg$_{0.2}$O), each containing 60 atoms in the supercell (Fig.~\ref{fig:str}). \\
From MO-11 to MO-14, the number of elements in the cationic site increases from 2 to 5. Absence of imaginary phonon frequencies in the phonon spectra confirms the dynamical stability of SQSs. Thus, dynamically stable SQSs truly reflect experimentally synthesized random solid solutions. In order to give more insight about entropy-stabilized MO phases, the thermodynamic properties, including enthalpy of mixing ($\Delta H_{mix}$) are
calculated using eq. (2). During $\Delta H_{mix}$ calculations, stable phases of constituent 1-cationic oxides are considered. Although MO solid solutions are stable in the rocksalt phase, their constituent oxides are not always stable in that particular phase. NiO, CoO and MgO are stable in the rocksalt phase, while CuO and ZnO are stable in the tenorite and wurtzite phases, respectively. According to eq. (2), if the oxide solid solutions possess negative $\Delta H_{mix}$, oxide solid solution phases can be considered as enthalpy-wise more favorable compared to their constituent oxide phases.

\begin{figure*}
    \centering
    \includegraphics[width=15cm]{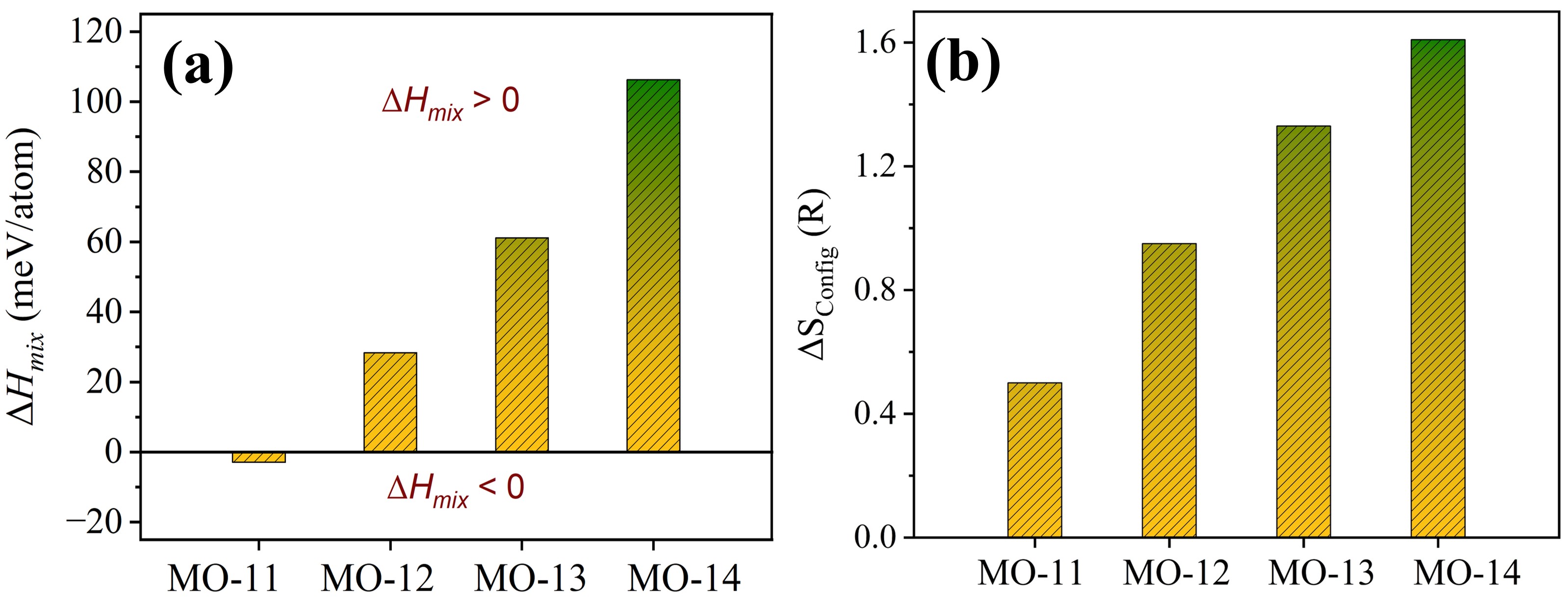}
    \caption{(a) Enthalpy of mixing ($\Delta H_{\mathrm{mix}}$) in units of meV/atom and (b) configurational entropy change in units of the universal gas constant, $R \approx 8.63\times10^{-5}\,\mathrm{eV\,K^{-1}\,atom^{-1}}$, for MO oxide solid-solution systems.}
    \label{fig:1}
\end{figure*}

As observed from Fig. \ref{fig:1}(a), MO-11 only exhibits low temperature stable phase. MO-12, MO-13 and MO-14 oxide solid solution phases exhibit positive $\Delta H_{mix}$. The positive $\Delta H_{\mathrm{mix}}$ values for MO-12, MO-13, and MO-14 are verified from both DFT and DFT+U frameworks (Table S1, Supporting Information). These positive $\Delta H_{\mathrm{mix}}$ values indicate that the formation of a homogeneous rock-salt solid solution is not enthalpically preferred with respect to the selected binary oxide reference phases.
This positive enthalpic contribution is also consistent with recent experimental thermodynamic measurements for the prototypical \textbf{(MgCoNiCuZn)$_{0.2}$O} high-entropy oxide. Sereda et al. experimentally determined an enthalpy of formation from the constituent binary oxides of 6.92$\pm$0.65 kJ/mol, corresponding to approximately 71.7$\pm$6.7 meV/atom.\cite{sereda} Thus, the experimentally measured formation enthalpy is positive, in agreement with the positive $\Delta H_{mix}$ obtained for our multicomponent systems and, in particular, the positive enthalpic penalty calculated for MO-14. Although the absolute values are not identical, the agreement in sign and magnitude supports the presence of a substantial enthalpic penalty associated with incorporating chemically distinct cations into the common rock-salt lattice.
This can be understood from the chemical and structural mismatch among the constituent cations. While NiO, CoO, and MgO naturally adopt rock-salt-type structures, CuO and ZnO prefer tenorite and wurtzite structures, respectively. Forcing these cations into a common rock-salt sublattice imposes an energetic penalty due to local strain, bonding mismatch, and deviations from their preferred coordination environments. Thus, the increasing positive $\Delta H_{\mathrm{mix}}$ from MO-12 to MO-14 reflects the increasing enthalpic cost of chemical complexity. In general, we can identify two different zones in the $\Delta H_{mix}$ plot. MO-11 belongs to a stable zone with $\Delta H_{mix}<0$. MO-12, MO-13 and MO-14 belong to the unstable zone with $\Delta H_{mix}> 0$. Thus, MO-12, MO-13 and MO-14 oxide solid solution phases cannot be considered as enthalpy-stabilized phases. As the number of elements at the cationic site increases from MO-11 to MO-14, a corresponding decrease in enthalpy-wise stability is observed. Negative (positive) values of $\Delta H_{mix}$ clearly indicate an exothermic (endothermic) reaction. Thus, the formation of the medium-entropy oxides MO-12 and MO-13 and the high-entropy oxide MO-14 involves the absorption of heat from external sources. In this work, “medium-entropy” denotes compositions with intermediate calculated configurational entropy and is not intended as a universal thermodynamic classification or evidence of phase stability. Similarly, an endothermic formation energy indicates that reactants are thermodynamically more favourable. The progressive increase in positive $\Delta H_{mix}$ from MO-12 to MO-14 indicates that the incorporation of multiple chemically and structurally dissimilar cations into a common rock-salt sublattice introduces increasing local strain and chemical mismatch. In particular, the coexistence of cations whose preferred binary oxides adopt different ground-state structures, such as tenorite CuO and wurtzite ZnO, imposes an energetic penalty when they are forced into the rock-salt lattice. Therefore, the high-temperature stability of these rock-salt oxide compositions cannot originate from enthalpic preference; rather, it must arise from entropy-driven lowering of the Gibbs free energy. Therefore, the stability of these multicomponent oxides cannot be inferred from $\Delta H_{\mathrm{mix}}$ alone. The positive mixing enthalpy of MO-12, MO-13, and MO-14 indicates that entropy must play a decisive role in the formation of stable single-phase rock-salt oxides at elevated temperatures. This provides the basis for analyzing the configurational, vibrational, and electronic entropy contributions in the following section.\\
\subsection{Entropy contribution}
As our concern is to identify high-temperature stable phases, it is necessary to analyze the role of entropy towards exploring the change in Gibbs free energy ($\Delta G_{mix}$) according to eq. (1). Change in Gibbs free energy not only involves $\Delta H_{mix}$ but also includes temperature multiplied by the entropy change ($\Delta S$). In fact, the entropy change includes contributions due to configurational degrees of freedom (DOF) ($\Delta S_{conf}$), vibrational DOF ($\Delta S_{vib}$), electronic DOF ($\Delta S_{elec}$) and magnetic DOF ($\Delta S_{mag}$). Configurational entropy is calculated based on fractional occupancy in the cationic site of MO as per eq. (3) [Fig. \ref{fig:1}(b)]. A random oxide material can be classified as a high-entropy oxide if it contains at least 5 or more elemental contributions in a single site ($S_{conf}\geq1.5R$). Based on eq. (3), MO-14 is classified as high-entropy oxide  followed by MO-13 and MO-12 as medium-entropy oxides [Fig.~\ref{fig:1}(b)]. This classification is based only on the magnitude of ideal configurational entropy. However, such a classification should not be directly equated with thermodynamic stability. A medium-entropy oxide may become entropy-stabilized if its positive $\Delta H_{\mathrm{mix}}$ is sufficiently compensated by the total entropy contribution at high temperature. Conversely, a high-entropy composition may still require a high stabilization temperature if the enthalpic penalty is large. Thus, $\Delta S_{\mathrm{conf}}$ is a useful descriptor of compositional disorder, but it is not a complete predictor of phase stability.\\
\begin{figure*}
    \centering
    \includegraphics[width=15cm]{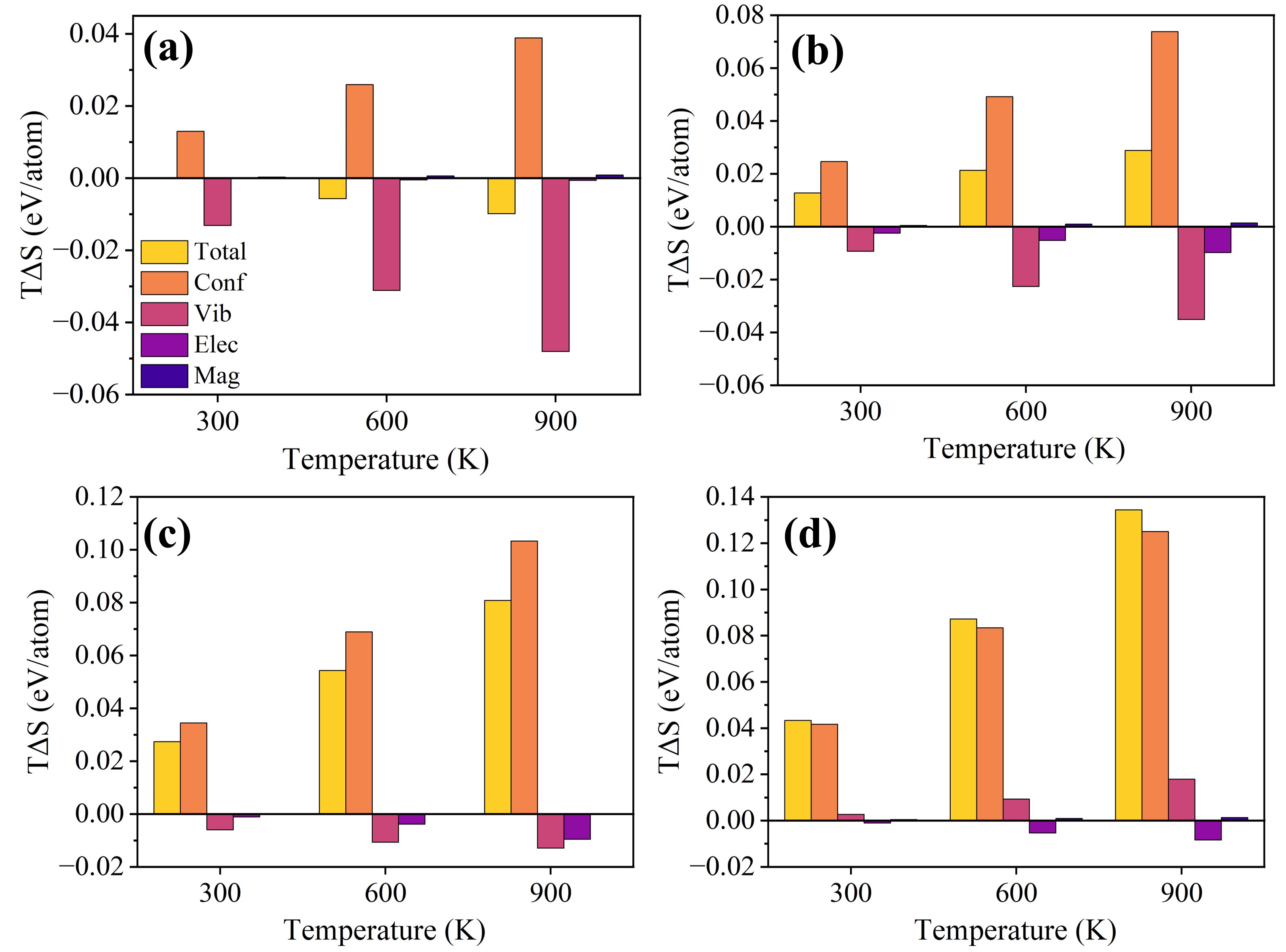}
    \caption{Temperature dependent entropy contributions at three different temperatures 300 K, 600 K, 900 K for (a) MO-11, (b) MO-12, (c) MO-13 and (d) MO-14.}
    \label{fig:2}
\end{figure*}
The comparative contributions of electronic, magnetic, vibrational and configurational entropy terms are shown in Fig. \ref{fig:2}. In case of MO-11, contributions from electronic and magnetic entropy terms are negligible compared to vibrational and configurational entropy contributions throughout the temperature regime. Thus, MO-11 exhibits a competition between the vibrational and configurational entropy terms. As around 300 K, these two terms contribute almost same, the total entropy term becomes negligible. However, as the temperature increases, corresponding vibrational entropy term contributes more, which results in the net nonzero $T\Delta S$. In all other oxides, from room temperature (300 K) onward, the total $T\Delta S$ term begins to contribute. In terms of magnetic entropy contributions, all MO's showcase positive contributions except for MO-13. Positive magnetic entropy in MO-11, MO-12, and MO-14 indicates that magnetic disorder or spin configurational freedom increases upon heating, favoring entropy-driven stabilization of these compositions. The negative magnetic entropy of MO-13 suggests reduced spin disorder or enhanced magnetic ordering, making its magnetic contribution unfavourable for stabilization at higher temperatures. Thus, MO-14 likely gains the greatest magnetic-entropy advantage because of its higher chemical and magnetic complexity. Starting from MO-12, the configurational entropy term starts dominating over vibrational, electronic, and magnetic entropy terms. The configurational entropy term becomes significant for 4- and 5-cationic oxides such as MO-13 and MO-14, respectively. Out of four oxide systems, MO-14 only possesses a positive contribution from the vibrational entropy term. The sign of $\Delta S_{\mathrm{vib}}$ provides insight into how lattice dynamics change upon mixing. A positive $\Delta S_{\mathrm{vib}}$ indicates that the multicomponent rock-salt phase has a larger vibrational entropy than the corresponding weighted binary oxide references, which may arise from phonon softening, mass disorder, and frustration in lattice geometry. This contribution further lowers $\Delta G_{\mathrm{mix}}$ at elevated temperatures. In contrast, a negative $\Delta S_{\mathrm{vib}}$ implies that the mixed phase is vibrationally less favorable than the reference oxides and therefore opposes entropy stabilization. Also, in case of 2- to 4-cationic oxides, the negative $\Delta S_{\mathrm{vib}}$ acts as a penalty that the configurational entropy has to overcome. However, with the transition from 4- to 5-cationic oxide system, $\Delta S_{\mathrm{vib}}$ acts synergistically with the configurational entropy working together to drastically lower the $\Delta G_{\mathrm{mix}}$ and stabilize the single-phase rock-salt structure. Hence, vibrational entropy is not universally stabilizing; its effect depends on the relative phonon spectra of the solid solution and the reference phases. As a result, the vibrational contribution, along with the significant configurational entropy term, makes the total $T\Delta S$ contribute largely to $\Delta G$. Recent experimental thermodynamic measurements by Sereda et al. reported a total entropy of formation of 15.95$\pm$0.16 J/mol-K at 298.15 K, corresponding to $T\Delta S\approx$0.0493 eV/atom.\cite{sereda} Our calculated value for MO-14 at 300 K is $\sim$0.043 eV/atom, showing reasonable agreement (within $\sim$13\%) with the experimental value and supporting the magnitude of the calculated total entropic contribution. The experimental thermodynamic data independently confirm the presence of a positive enthalpic penalty and a substantial entropic contribution, both of which are central to the thermodynamic picture developed in our calculations.

\subsection{Gibbs free energy of solid solutions}    
\begin{figure}[t] 
    \centering
    \includegraphics[width=10cm]{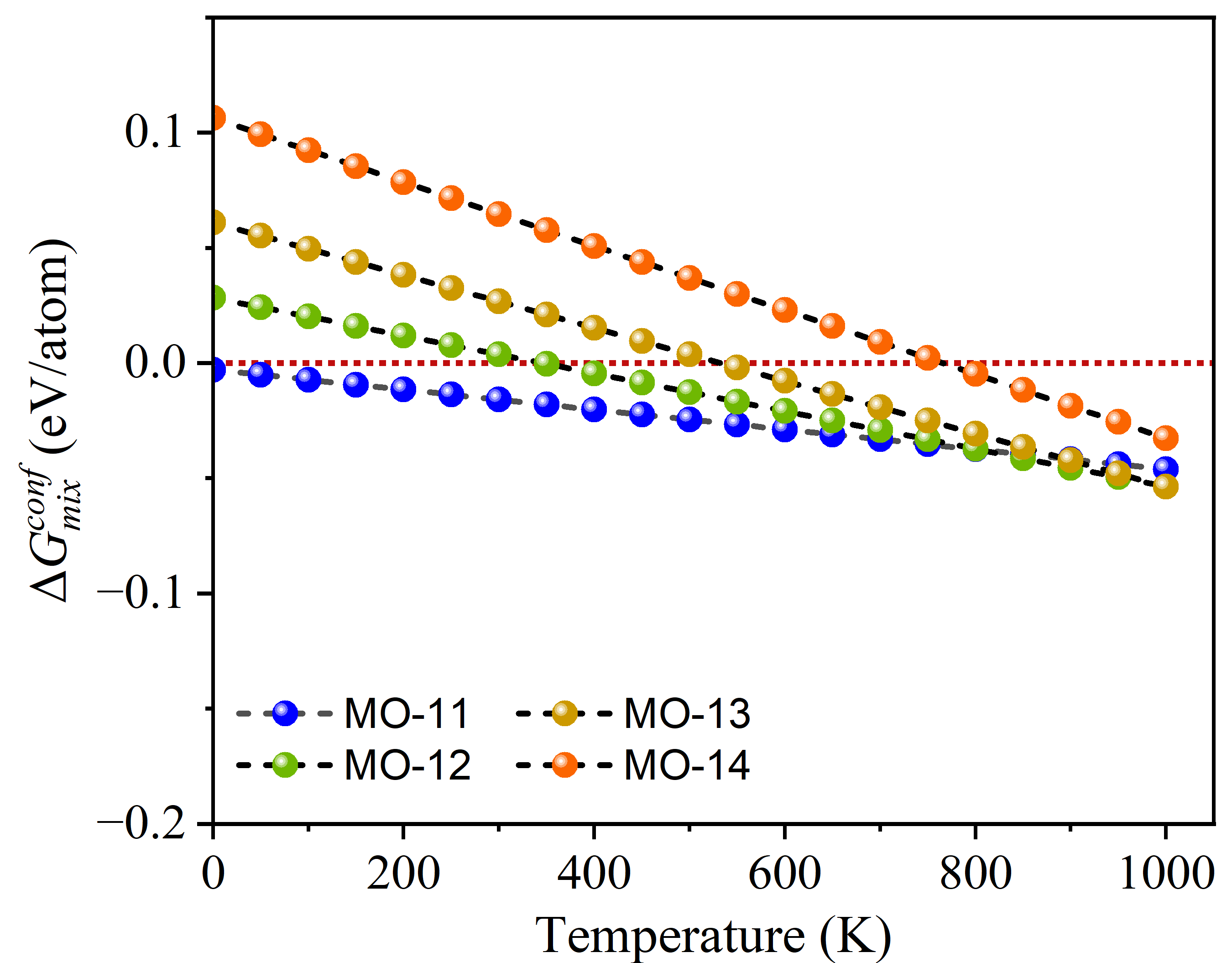}
    \caption{Variation of Gibbs free energy change with temperature considering only configurational entropy change. Different MO systems are represented by different colors as indicated in the inset.}
    \label{fig:3}
\end{figure}

\begin{figure*}
    \centering
    \includegraphics[width=15cm]{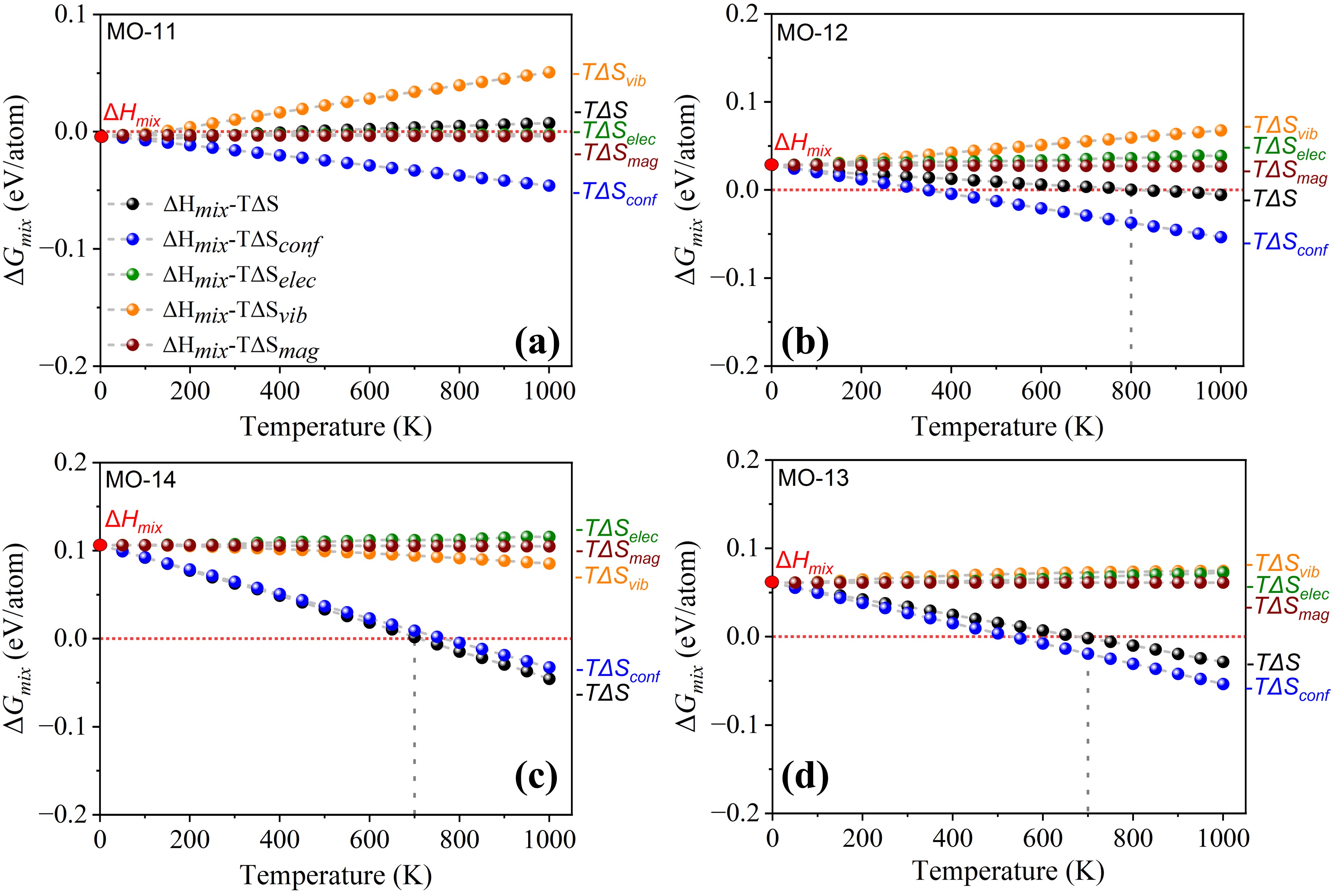}
\caption{Variations of Gibbs free energy change due to mixing as a function of temperature for (a) MO-11, (b) MO-12, (c) MO-13 and (d) MO-14 systems. Corresponding individual contributions are represented by different colors as indicated in the inset of the figure.}
    \label{fig:4}
    \end{figure*}

Taking into consideration only the configurational entropy term, change in Gibbs free energy can be expressed as follows.
    \begin{equation}
        \Delta G_{mix}^{conf} = \Delta H_{mix} - T\Delta S_{conf}
    \end{equation}
Fig. \ref{fig:3} depicts the variation $\Delta G_{mix}^{conf}$ as a function of temperature according to eq. 5. 
At 0 K, MO-12, MO-13, and MO-14 all have a positive enthalpy of mixing. However, when configurational entropy is taken into account, $\Delta G_{mix}^{conf}$ changes from positive to negative values at around 350 K (MO-12), 550 K (MO-13) and 750 K (MO-14), respectively. As the temperature increases further, MO-12 and MO-13 attain more negative $\Delta G_{mix}^{conf}$ with a value of $> 50$ meV/atom at 1000 K temperature. The inclusion of the configurational entropy term, $T\Delta S_{conf}$ does not guarantee the high-temperature stability of these oxides. The $\Delta G_{\mathrm{mix}}^{\mathrm{conf}}$ analysis demonstrates the stabilizing tendency of configurational entropy, but it also highlights the limitation of using $\Delta S_{\mathrm{conf}}$ alone. Since the actual free energy contains vibrational and electronic entropy contributions, a stability criterion based only on $\Delta S_{\mathrm{conf}}$ may either overestimate or underestimate the stabilization temperature. Therefore, the total Gibbs free energy including $\Delta S_{\mathrm{conf}}$, $\Delta S_{\mathrm{vib}}$, $\Delta S_{\mathrm{elec}}$ and $\Delta S_{\mathrm{mag}}$ provides a more reliable thermodynamic description. To give more insight about high-temperature stability of these oxides, it is necessary to include other possible contributions towards the entropy change.
In fact, taking into consideration vibrational, electronic, and magnetic contributions, the change in Gibbs free energy of mixing can be written as follows.
    \begin{equation}
        \Delta G_{mix} = \Delta H_{mix} - T\Delta S_{conf}- T\Delta S_{vib}- T\Delta S_{elec} - T\Delta S_{mag}
    \end{equation}
Here $T\Delta S_{vib}$, $T\Delta S_{elec}$ and $T\Delta S_{mag}$ represent entropy change due to lattice vibrations, electronic and magnetic contributions respectively. Fig. \ref{fig:4}(a) depicts the variations of $\Delta G_{mix}$ with temperature including all possible entropic contributions for MO-11. In case of MO-11, electronic and magnetic entropy contributions are almost similar. However, with increasing cationic occupancy such as for MO-12, MO-13 and MO-14, a little deviation between $\Delta G_{mix}^{elec}$ and $\Delta G_{mix}^{mag}$ is observed.
A very small difference between $\Delta G_{mix}$ and $\Delta G_{mix}^{elec}$ is observed around high-temperature, which basically arises due to the competition between vibrational and configurational entropy contributions. Around room temperature, both the contributions are similar. However, with increasing temperature, the vibrational term contributes more in the negative regime, in such a way that $\Delta G_{mix}$ becomes positive even at a high-temperature of 1000 K. Thus with all possible entropy consideration, MO-11 becomes an unstable phase at high-temperature regime. A different scenario is observed in case of other MO systems. MO-12 possesses a positive $\Delta H_{mix}$ at 0 K temperature. However with increasing temperature as the entropy terms are included, $\Delta G_{mix}$ goes through a transition from positive to negative values at a temperature of 800 K. Around both low and high-temperature regime, the configurational contribution becomes twice that of the vibrational contribution [Fig. \ref{fig:4}(b)]. Around 1000 K temperature, considering all entropy contributions, MO-12 possesses a negative $\Delta G_{mix}$ of 5.61 meV/atom. Thus, contrary to enthalpy-stabilized MO-11, MO-12 exhibits a thermodynamically favorable $\Delta G_{mix}$ relative to the selected binary-oxide reference states at elevated temperature. MO-13 exhibits negative values of both $T\Delta S_{vib}$, $T\Delta S_{elec}$ and $T\Delta S_{mag}$ throughout the temperature regime. Around high temperatures, the vibrational entropy contribution is quite comparable with the electronic entropy contribution. However, even at high temperatures, the magnetic entropy contribution is very small compared to the vibrational and electronic entropy terms. Thus, considering the vibrational and electronic entropy contributions, $\Delta G_{mix}$ shows more positive values compared to $\Delta H_{mix}$ at higher temperature. However, significant contribution from configurational entropy makes transition of the $\Delta G_{mix}$ from positive to negative at $T\geq 700$ K [Fig.~\ref{fig:4}(c)]. Thus, 3- and 4-cationic oxide systems such as MO-12 and MO-13 can possess significant configurational entropy which can promote entropy-assisted stabilization relative to the selected reference states. On the other hand, because of high positive $\Delta H_{mix}$, MO-14 is considered as enthalpy-wise unstable phase as 0 K. In addition, being a 5-cationic oxide system, MO-14 exhibits the highest configurational entropy (1.61\,R) contribution among all oxide systems considered in this study. Because of that main contribution in $T\Delta S$ comes from configurational entropy. As the temperature increases, vibrational term $T\Delta S_{vib}$ also contributes significantly in $T\Delta S$ [Fig. \ref{fig:4}(d)]. In the high-temperature regime, the large positive contributions of $T\Delta S_{conf}$ and $T\Delta S_{vib}$ projects $T\Delta S$ to a significant quantity. This $T\Delta S$ is sufficient enough to offer a transition of $\Delta G_{mix}$ from positive to negative values around 700 K. Further, around 1000 K temperature, $\Delta G_{mix}$ reaches a more negative value of -45.7 meV/atom. The predicted entropy-driven stabilization of MO-14 is qualitatively consistent with experimental observations of (Mg,Ni,Co,Cu,Zn)O, which show temperature-dependent structural evolution, segregation, and reversible phase transformations.\cite{coduri,dupay} These results highlight the roles of competing phases and kinetic effects in addition to thermodynamic driving forces. The calculated full-entropy crossover temperatures for MO-12, MO-13, and MO-14 are approximately 800 K, 700 K, and 700 K respectively. Related multicomponent rock-salt oxides are generally synthesized or annealed at higher temperatures, typically around 850–1000$^{\circ}\mathrm{C}$. The difference is expected because the calculated crossover represents the point at which $\Delta G_{mix}$ changes sign relative to the selected reference oxides, whereas experimental formation also involves kinetic limitations, oxygen chemical potential, precursor effects, and competing phases. Thus, these temperatures should be regarded as idealized thermodynamic indicators rather than exact synthesis temperatures. As MO-11 exhibits positive $\Delta G_{mix}$, it cannot be considered as stable around high-temperature. On the other hand, although having high positive $\Delta H_{mix}$, MO-14 exhibit significant negative $\Delta G_{mix}$ around high-temperature because of high configurational entropy. MO-14 exhibits entropy-assisted thermodynamic stabilization relative to the selected reference states. Even MO-12 and MO-13 having medium configurational entropy, can also offer high-temperature entropy-stabilized and thermodynamically stable phases. Overall, these results demonstrate that entropy stabilization in rock-salt oxides should not be reduced to a fixed configurational entropy threshold. The conventional $\Delta S_{\mathrm{conf}} \geq 1.5R$ criterion is useful for classifying high-entropy compositions, but it does not by itself establish entropy-stabilized phase formation. The formation of a single-phase rock-salt oxide is governed by the balance between the enthalpic penalty associated with incorporating chemically distinct cations into a common lattice and the total entropy gain available at elevated temperature. In this framework, MO-12 and MO-13 are particularly important because they show that medium-entropy compositions can become stable when their moderate positive $\Delta H_{\mathrm{mix}}$ is compensated by configurational and non-configurational entropy terms. \cite{fracchia2022configurational,he2024development,aamlid2023understanding} MO-14, although having the largest $\Delta S_{\mathrm{conf}}$, also possesses the largest enthalpic penalty, showing that high-entropy and high stability are not necessarily equivalent. This conclusion is consistent with recent experimental reports on quaternary medium-entropy rock-salt oxides.\cite{he2024development} For example, single-phase $(\mathrm{Mg}_{0.25}\mathrm{Co}_{0.25}\mathrm{Ni}_{0.25}\mathrm{Zn}_{0.25})\mathrm{O}$ has been synthesized in the $1000$--$1200\,^{\circ}\mathrm{C}$ temperature range and shown to possess good thermal stability. Such studies support the idea that the five-cation high-entropy condition is not a strict requirement for stabilizing rock-salt oxide solid solutions.\cite{aamlid2023understanding, manchon2025descriptors, pitike2020predicting} The present calculations provide a thermodynamic basis for this observation by showing that medium-entropy compositions can also reach negative $\Delta G_{\mathrm{mix}}$ at elevated temperature and are in line with the recent observation made experimentally\cite{kumar2025configurational}.\\
Although inclusion of Hubbard $U$ in the methodology to treat localized 3$d$ electrons, modifies the absolute mixing enthalpies, the relative thermodynamic trends and entropy-driven crossover behavior remain unchanged. The present analysis considers selected binary oxide reference phases; however, competing phases, local ordering, oxygen non-stoichiometry, and cation valence may also affect phase stability. Experiments on \textbf{(MgCoNiCuZn)$_{0.2}$O} show high-temperature Cu loss and Zn volatilization, reducing configurational entropy and promoting multiphase formation.\cite{Web} Furthermore, the experimentally observed room-temperature phase assemblage may depend strongly on synthesis temperature, annealing time, cooling rate, and cation diffusion kinetics.\cite{spurling2022entropy, aamlid2023understanding, manchon2025descriptors, pitike2020predicting} Thus, $\Delta G_{\mathrm{mix}} < 0$ should be regarded as a thermodynamic indicator rather than an exact synthesis or phase-boundary prediction.\\


\section{Conclusion}
In conclusion, using a first-principles-based thermodynamic framework, we systematically investigated the temperature-dependent phase stability of rock-salt oxide solid solutions containing two to five principal cation components, denoted as MO-11 to MO-14. MO-11 exhibits a negative $\Delta H_{\mathrm{mix}}$ at 0 K, whereas MO-12, MO-13, and MO-14 have positive $\Delta H_{\mathrm{mix}}$, indicating that their homogeneous rock-salt solid-solution formation is not enthalpically favoured with respect to the selected binary oxide reference phases. The Gibbs free-energy analysis shows that this 0 K picture changes at finite temperature: MO-11 becomes thermodynamically unfavorable at elevated temperature, whereas MO-12, MO-13, and MO-14 are stabilized by the increasing entropy contribution. Among the entropy terms, vibrational entropy dominates for MO-11, while configurational entropy provides the major stabilizing contribution for MO-12, MO-13, and MO-14, with additional contributions from vibrational, electronic and magnetic entropy. Importantly, MO-14 has the highest configurational entropy ($\Delta S_{\mathrm{conf}}\approx$1.61R) but also the largest positive $\Delta H_{\mathrm{mix}}$, demonstrating that configurational entropy alone does not determine thermodynamic stability. In contrast, the lower-entropy MO-12 and MO-13 systems can achieve negative $\Delta G_{\mathrm{mix}}$ at elevated temperatures because their smaller enthalpic penalties are compensated by the total entropy contributions. Thus, the thermodynamic stability of compositionally complex rock-salt oxides should be assessed from the full $\Delta G_{\mathrm{mix}}$ balance rather than from the number of cation components or the $\Delta S_{\mathrm{conf}}\geq$1.5R criterion alone. The comparison with recent experimental thermodynamic data for (MgCoNiCuZn)$_{0.2}$O further supports this thermodynamic picture, with the experimentally reported positive formation enthalpy and substantial entropic contribution being consistent with the calculated enthalpic penalty and entropy-driven stabilization. Overall, these results distinguish compositional complexity from genuine finite-temperature stabilization and show that medium-entropy compositions can also be thermodynamically stabilized under appropriate conditions. The present analysis is based on decomposition into selected binary oxide reference phases and does not explicitly account for kinetic effects, oxygen non-stoichiometry, cation short-range ordering, magnetic entropy, defect formation, or competing ternary/intermediate phases. Therefore, the predicted stabilization temperatures should be regarded as idealized thermodynamic indicators rather than exact experimental phase or synthesis boundaries.

\section{Declaration of competing interest}
\footnotesize
The authors declare that they have no known competing financial interests or personal relationships that could have appeared to influence the work reported in the paper.

\section{Author Contribution Statement}
\textbf{SG} Conceptualization, Simulation, Formal analysis, Writing original draft. \textbf{AK}: Conceptualization, Formal Analysis, Writing-review and editing. \textbf{SKP}: Conceptualization, Discussions, Supervision, Finalizing manuscript.
\section{Acknowledgment}
 SG sincerely acknowledges ANRF, Govt. of India (File number: PDF/2025/001037) for providing NPDF fellowship. SKP acknowledges ANRF JC Bose Fellowship and ANRF, Govt. of India for the Financial support. SG and SKP would like to thank the National Supercomputing Mission, PARAM Yukti facilities for providing computational resources.

\bibliographystyle{elsarticle-num}
\bibliography{ref}
\end{document}